# VOLTAGE AND TEMPERATURE DEPENDENCE OF HIGH-FIELD MAGNETORESISTANCE IN ARRAYS OF MAGNETIC NANOPARTICLES

Reasmey P. Tan, Julian Carrey and Marc Respaud

Université de Toulouse; INSA,UPS; LPCNO, 135 avenue de Rangueil, F-31077 Toulouse, France
CNRS; LPCNO, F-31077 Toulouse, France

## Abstract :

Huge values of high field magnetoresistance have been recently reported in large arrays of CoFe nanoparticles embedded in an organic insulating lattice in the Coulomb blockade regime. An unusual exponential decrease of magnetoresistance with increasing voltage was observed, as well as a characteristic scaling of the magnetoresistance amplitude versus the field-temperature ratio. We propose a model which takes into account the influence of paramagnetic impurities on the transport properties of the system to describe these features. It is assumed that the non-collinearity between the core spins inside the nanoparticles and the paramagnetic impurities can be modelled by an effective tunnel barrier, the height of which depends on the relative angle between the magnetization of both kind of spins. The influence on the magnetotransport properties of the height and the thickness of the effective tunnel barrier of the magnetic moment of the impurity, as well as the bias voltage are studied. This model allows us to reproduce the large magnetoresistance magnitude observed and its strong voltage dependence, with realistic parameters.

## I. INTRODUCTION

When metallic nanoparticles (NPs) are embedded in an insulating matrix, the tunnelling between the NPs controls the electronic transport mechanisms.[1,2] In this case, an energy $E_C \propto 1/d$ where $d$ is the NP diameter is required to charge a NP. At a temperature $T$ well below this energy, conduction through the NPs is only possible above a threshold voltage.[3] This regime is referred as Coulomb blockade regime. Besides, if the NPs are magnetic and spin-polarised, tunnel rate transmission is influenced by the relative orientation of their magnetic moments.[4-6] This phenomenon leads to the well-known tunnel magnetoresistance (MR) effect : the resistance $R$ of the array changes from a minimum value when the magnetic moments are parallel to a maximum value when the magnetic moments are in a disordered state. Thus, the low state of resistance ($R_{min}$) is obtained for an external field applied above the saturation field ($H_{sat}$) of the NPs assembly and the high state of resistivity ($R_{max}$) for an external field equal to the coercive field, as the magnetic moments of the NPs are randomly oriented.

Relevant models have been proposed to explain tunnel MR in NPs arrays.[7,8] They have been essentially focused on the competition between Coulomb blockade and magnetic energy of the granular assemblies. Particularly, in a model of non-interacting NPs based on MR in magnetic tunnel junctions,[9] Inoue and Maekawa[8] argued that the amplitude of the tunnel MR is determined by the spin polarisation $P$ of the NPs:

$$\text{MR} = \frac{R_{max} - R_{min}}{R_{min}} = \frac{2P^2}{1+P^2} \tag{1}$$

The field dependence of the resistance ($R(H)$) can also be expressed as a function of the normalised magnetisation $m = M/Ms$ of the assembly, where $M$ is the magnetisation and $Ms$ the saturated magnetisation of the system :

$$R(H) \propto \frac{m^2(H) P^2}{1+m^2(H) P^2} \tag{2}$$

In this model, the MR variation as a function of temperature [MR($T$)] is due to the temperature dependence of the magnetisation and of the spin polarisation. Moreover, size distribution of NPs is neglected since tunnelling events only occurs between the largest NPs with identical diameter. Mitani and *al.*[10] studied the effect of co-tunneling effects between large grains via smaller ones. Their calculations show that a *MR* enhancement in the Coulomb blockade regime is expected but is still limited to few ten percents. Besides, the voltage dependence of the MR was found to be negligible (compared to the *T*-dependence) since large number of NPs are involved in transport mechanism in NPs arrays.

Despite their accuracy, two main features of MR measurements in NPs assemblies are not explained by these models:

i/ according to (1), the maximum value of MR in NPs assemblies is 50 % assuming a full spin polarisation. This could not explain the large values measured by Chen *et al.*[11] up to 158 % at room temperature and more than 1000 % at low temperature in polycrystalline $Zn_{0.41}Fe_{2.59}O_4$ grains separated by α-$Fe_2O_3$ grain boundaries. The strong MR amplitude has been explained by magnetic correlations due to relative orientation between the magnetisations of the grains and the grain boundaries.

ii/ according to (2), when $m = 1$ (saturated magnetisation), $R(H_{sat})$ is supposed to reach the value of $R_{min}$. However, unexpected high-field MR has been observed in different systems such as granular films,[12,13] NPs arrays,[14-17] or polycrystalline films with grain boundaries.[18,19] In these experiments, magnetoresistance curves [MR*(H)*] are unsaturated even in large fields $H \gg H_{sat}$. This high-field behaviour has been attributed to the influence of paramagnetic impurities dispersed in the insulating matrix[12,13,17] or to spin disorder at the surface of NPs.[13-19]

Few models have been proposed to explain these anomalous MR behaviours. Among them, the one stated by Holdenried and *al.*[20] take into account a spin disorder at the surface of NPs which depends explicitly on the temperature. As *T* is increased, the spin disorder increases and reduces the MR amplitude. Thus, this model only proposes an alternative interpretation to MR*(T)* dependence in Mitani and *al.*'s[10] model, but does not explain MR values greater than 50 %. Another description of spin disorder in magnetic NPs has been proposed by Huang and *al.*.[21] In their model, a NP is described as a ferromagnetic core surrounded by a layer of disordered spins. It is assumed that the canted spins act as an additional effective tunnel barrier in serie with the insulating layer when no magnetic field is applied. By applying a magnetic field, the reduction of the canting of the disordered shell with respect to the ferromagnetic core induces a collapse of the additional tunnel barrier with a concomitant large MR. This model is able to explain quantitatively the huge amplitude of MR observed experimentally by Chen *et al.*[11] and its temperature dependence. Up to now, the voltage dependence of the MR was not addressed neither experimentally nor theoretically.

We recently reported the observation of huge values of high-field MR in arrays of CoFe NPs separated by insulating thin organic layers. We showed that the amplitude of the MR only depends on the *H/T* ratio, and strongly varies with the applied voltage, with in some cases, an exponential increase when decreasing voltage. We attributed this unusual behaviour to the presence of paramagnetic impurities in the sample.[17] In the present article, we show that a simple model is able to reproduce the features of this MR. The current through a NP and its neighbouring impurity is calculated on the basis of Simmons and Fowler Nordheim's models. Similarly to Huang and *al*.,[21] an effective tunnel barrier due to the presence of the paramagnetic impurities is considered. We demonstrate that by adjusting the height and the thickness of the effective barrier, the Fowler Nordheim's model can well reproduce the experimental features, namely the field, temperature and bias voltage influence on the MR properties.

The paper is organised as follows. First, we recall the main experimental results obtained on three-dimensional super-lattices of CoFe NPs (section **II**). To analyze these experimental properties, we propose a simple model of transport through the effective tunnel barrier created by the paramagnetic impurities. The influence of the various parameters is studied in section **III**. In section **IV**, we compare the numerical results with the experimental data. Finally, we discuss the parameters obtained from the fits as well as the validity of the model in section **V**.

## II. EXPERIMENTAL RESULTS

First, we recall briefly the magnetotransport properties of 3D millimetre-long super-lattice of CoFe NPs.[17] The super-lattices are chemically synthesized using organometallic decomposition in mild conditions.[22] The system is composed of spherical CoFe NPs with mean diameter $D \sim 15$ nm separated by thin insulating barrier $L \sim 2$nm which are composed of organic ligands (mixture of long chain amine and carboxylic acids). The system is ferromagnetic up to room temperature. An original mechanism of MR due to collective effect related to Coulomb blockade was observed on the super-lattices.[23] In this paper, we only focus on the high-field MR measurements.

In summary, significant high-field MR was only found for temperature ranging from 2 K to 10 K (see Fig.1). At $T = 3.15$ K, MR measurements were performed at different voltages, showing huge value of MR > 3000 % at $\mu_0 H = 8.8$ T and $V = 20$ V while 40 % is obtained for bias voltage $V = 200$ V (see Fig.1c). Besides, the shape of the MR curve is different depending on the MR amplitude: a linear dependence for low MR, and a exponential one for larger MR. The complete voltage dependence was deduced at $T = 2.75$ K (see. Fig.1d) from the measurement of two $I(V)$ curves without [$I(0$ T$,V)$] and under a magnetic field $\mu_0 H = 8.8$ T [$I(8.8$ T$,V)$] according to the expression :

$$\mathrm{MR}(V) = (I(8.8 \text{ T},V) - I(0 \text{ T},V)) / I(0 \text{ T}, V). \tag{3}$$

The MR($V$) curve displays an exponential decrease as $V$ is increased. The lack of points in the MR($V$) curve at low voltage is due to noise. Another interesting property observed in the range of temperature between 2 K and 10 K is shown in Fig.1a. For a given bias voltage ($V = 200$ V), all MR curves superpose on an universal MR curve when they are plotted as a function of the $H/T$ ratio.

We will attempt to explain the experimental behaviour of the MR amplitude, especially its huge value, its $H/T$ and voltage dependence and the change of slope in $R(H)$ characteristics.

## III. MODEL

In this part, we present the model used to describe the effect on transport properties of magnetic impurities located in the organic barriers. We assume that the individual magnetic moment creates an effective tunnel barrier which is progressively removed by applying an external magnetic field (see.Fig.2), *i-e*, by aligning the disordered spins with the ferromagnetic grains. As suggested by Huang and *al*.,[21] the misalignment of the spin of the magnetic impurity ($\boldsymbol{\mu}_j$) with the magnetic moment of the NP ($\boldsymbol{\mu}_i$) leads to an effective tunnel barrier $\phi$,

$$\phi = J\,(1 - \langle\boldsymbol{\mu}_i.\boldsymbol{\mu}_j\rangle) \tag{4}$$

where $J$ is the height of the effective barrier. $\boldsymbol{\mu}_i$ is supposed to be fixed and saturated for weak magnetic field. Thus, $\langle\boldsymbol{\mu}_i.\boldsymbol{\mu}_j\rangle$ is given by a simple Langevin function [$L(\xi) = \coth(\xi) - 1/\xi$, with $\xi = \mu_j H / k_B T$ ]. This leads to,

$$\phi = J\,(1 - L(\xi)) \tag{5}$$

The tunnel current density through the effective barrier is calculated considering two models of tunnel current density: Simmons[24] and Fowler-Nordheim's [25] models. In both cases, the tunnel current depends exponentially on the height $\phi$ and the thickness $s$ of the barrier, but with complementary domains of validity.

Simmons derived the current-voltage [$I(V)$] expression in the case of rectangular barrier,

$$I_S \propto (\phi - eV/2)\exp(-A.s(\phi - eV/2)^{1/2}) - (\phi + eV/2)\exp(-A.s(\phi + eV/2)^{1/2}). \quad (6)$$

where $A = 4\pi/h(2m_e e)^{1/2}$, and $h$ is the Planck's constant, $m_e$ the mass of an electron. It should be noticed that this expression has been derived in the case of weak polarisation since $eV < \phi$. On the opposite, the Fowler-Nordheim's current $I_F(V)$ corresponds to the case of strong polarisation $eV > \phi$, with,

$$I_F \propto V^2/(\phi.s^2).\exp(-A.\,s.\,\phi^{3/2}/V). \quad (7)$$

In our model, we consider that the tunnelling process from the NP to the impurity controls the total current flow and thus the total MR. Direct spin dependent tunnelling between two NPs and the effect of Coulomb blockade on the current are both neglected. Moreover, the resistance of the insulating layer is supposed to be independent of the effect of bias voltage. So it is considered as a constant contribution $R_T$ and is simply added to the total resistivity. Since we only study the case of half a barrier (a NP and an impurity), $R_S = R_T/2$ is introduced. Consequently, the MR ratio becomes,

$$MR = \frac{(R_{max} + R_S) - (R_{min} + R_S)}{(R_{min} + R_S)} \quad (8)$$

In the next part, we first assume $R_S = 0$. The influence of $R_S$ on the MR properties will be shown later.

We now present the numerical calculations obtained from this model. Fig.3a and 3b show characterictic $I(V)$ curves calculated with the expressions of Simmons and Fowler Nordheim for typical parameters of tunnel barriers and a ratio $\mu/T = 0.1\ \mu_B.K^{-1}$. In both cases (Simmons and Fowler-Nordheim), the $I(V)$ curves exhibit higher conductance when applying a magnetic field, as a consequence of the decrease of $\phi$. Attention should be paid to the domain of validity of both models, since they are defined in a range of voltage values depending on $\phi$. Particularly, $I_S$ is valid for $eV < \phi$ and since $\phi(\mu_0 H) < J$, the domain is limited to $eV < \phi(\mu_0 H)$. In the other case, the domain of validity for Fowler-Nordheim remains the same $eV > J$. These two conditions are depicted in Fig.3 by two vertical dash lines.

The voltage dependence of MR($V$) (Fig.3c and 3d) for various set of barriers parameters (height/thickness) are calculated from the $I(V)$ characteristics of Fig.3a and 3b. MR amplitude in Simmons model saturates at low voltage. This behaviour is easily explained by the fact that *I(V)* characteristics are linear for very small voltage values.[24] When *V* is increased, MR ratio increases. In the case of Fowler-Nordheim's current, the influence is opposite to the previous one : MR(*V*) strongly decreases when *V* is increased. As a result, maxima values of MR are obtained at the crossover of both models, when e*V* is of the order of $\phi$.

Fig.4a and 4b display *R(H)* curves for each tunnel current model at various voltages for $\mu/T = 0.3$ $\mu_B.K^{-1}$, *s* = 1 nm and *J* = 1 V. Similarly to Fig.3c, Simmons model leads to identical *R(H)* dependences for low voltages, with an amplitude increasing slightly when increasing voltage. Fowler-Nordheim's expression leads to a continuous drop of MR when increasing *V*. Besides, *R(H)* shapes depend on the applied voltage, leading to two characteristic MR dependences. Fig.4b illustrates these behaviours, with a linear *R(H)* for *V*~ 10$\phi$ and a strongly non linear characteristic for *V*~ $\phi$.

We now explore the influence of $R_S$. According to the assumption made on $R_S$ (independent of the voltage), $R_S$ only affects the MR amplitude for strong value of *V*. Fig.4c illustrates this behaviour: $R_S$ reduces the MR amplitude when *V* increases.

To complete this theoretical study, we systematically investigate the influence of the barrier parameters and temperature with $R_S$ = 0. The MR has been calculated at a given magnetic field of 5T. Fig 5a and 5b display the calculations for $\mu/T = 0.1$ $\mu_B.K^{-1}$, as a function of *s* and *J*, for *V* = 0.05 V in the case of Simmons current and *V* = 2.5 V for Fowler-Nordheim current. These values of *V* are chosen in order to be in the range of validity of each expression. For thick and high barriers, very large amplitudes of MR are obtained. We emphasize the fact that for a given value of MR, several combinations of parameters (*s,J*) are possible.

We now investigate the effect of the temperature on the MR magnitude. Fig.5c and 5d show a similar study to Fig.5a and 5b, but for a ratio $\mu/T = 0.3$ $\mu_B.K^{-1}$, *i-e* at lower temperature. These results show the strong dependence of the MR ratio on temperature since it could reach more than 10 000 % for a thick and high barrier. For the same thickness and height of the barrier, the MR value is about 500 % for $\mu/T = 0.1$ $\mu_B.K^{-1}$.

In summary, high value of MR are obtained in Fowler-Nordheim and Simons cases: i/ at low temperature, ii/ for *V*~ $\phi$, iii/ for large values of *s* and/or *J*. However, the voltage dependence of MR is opposite in the two models, and only the Fowler-Nordheim's expressions lead to an decrease of the MR amplitude with increasing voltage.

## IV. COMPARISON WITH EXPERIMENTS

We now use our model to fit the experimental results obtained in $R(H)$ and MR$(T)$ characteristics. First, it is primordial to specify the meaning of the voltage applied in experiments. Actually, the millimetre-long CoFe super-crystals were connected using gold wires and silver painting to make the contacts. This leads to a typical distance of 0.1 mm between the contacts. With $D \sim 15$ nm and $L \sim 2$ nm, thousands particles in series and in parallel are measured. Thus, if $V = 20$ V is applied on the whole super-lattice, $V \sim 3$ mV is expected for one particle. However, voltage distribution in NPs arrays is quite complex and hard to predict. Imamura and *al.*[26] pointed out the fact that NPs assemblies modelled as tunnel junctions show current paths which depend on the value of the NPs capacitance. As a result, the bias voltage is not proportionally distributed between all particles. Considering these issues, we consider the voltage drop applied on one particle as a free parameter in our fits. Let $V_{min}$ be the voltage value used in the calculations and equivalent to $V = 20$ V applied in MR measurements. To keep the trend of voltage effect on MR curves, the ratio between the three applied voltages (20, 70 and 200 V) are conserved in the numerical calculations.

Experimentally, we observed that increasing $V$ (see Fig.1d) causes an exponential decrease of the MR$(V)$ curve. This result indicates that the experimental $R(H)$ characteristics (see Fig.1c) can be reproduced only if Fowler-Nordheim's model is considered (see Fig.3c, 3d and 4b). Attention should be paid to the fact that our model gives us several set of parameters which can fit satisfactorily the experimental curves, including meaningless cases such as too large barrier thicknesses. Thus, we had to reduce some range of values for the parameters. In particular, the barrier thickness $s$ has been restricted to the range 0-2 nm. We also assumed that impurities are isolated atoms, so only multiples of $S = 1/2$ were taken as possible values for $\mu_j$. $s$, $J$, $V_{min}$ and $R_S$ are set as free parameters. Best fits of $R(H)$ curves at $T = 3.15$ K (Fig.6b, 6c and 6d) are then obtained with $\mu_j = 1$ $\mu_B$, $s = 0.5$ nm, $J = 2$ V, $V_{min} = 2.2$ V and $R_S = 15$ k$\Omega$.

Experimental $R(H)$ curves are well reproduced by our simulations even if there is a small discrepancy for $V = 70$ V. The important result is that voltage dependence on the MR amplitude is reproduced as well as the influence on the shape of the curves. Fig.6a shows the calculated MR$(T)$ with the same fitting parameters. The experimental decrease with temperature is reproduced, with the absence of MR above 10 K and the growth of MR at lower temperature.

## V. DISCUSSION

We now discuss the validity of the model. The numerical results obtained are based on a basic picture and the model could be improved to give a more accurate description of the high-field MR. First, for simplicity, we chose the Langevin function to describe the drop of the effective barrier. By doing this, we implicitly assumed that the NPs are surrounded by an homogeneous barrier, the height of which depends on θ, the average angle between a spin located on the impurity and the core. A more correct vision of the issue would require to consider an inhomogeneous tunnel barrier and to calculate all the values of tunnel current as a function of all possible θ orientations. Then the total tunnel current would be the sum of the individual tunnel currents weighted by the magnetic field-dependent probability to find a impurity spin making an angle θ with respect to the core. This probability has also to be considered instead of the Langevin function because this latter is only valid in the classical limits. Second, we simply calculated the tunnel current using Simmons and Fowler-Nordheim's equations. It has been shown that other models lead to a better description of the spin-dependant transport through a tunnel barrier, especially when dealing with the voltage dependence of the magnetoresistance.[27,28] Finally, the MR is calculated on one particle while a large number of particles are measured. Nevertheless, our approach appears sufficient to explain most of the experimental observations.

Another point has to be emphasized. In general, canted spins at the surface of NPs or grain boundaries have been invoked to explain MR properties of oxides NPs.[14-16,18,19] In our case, FeCo NPs are metallic. Thus, unless an adventitious oxidation, the presence of paramagnetic species at the surface would be due to a surface state modified by organic ligands such like carboxylic chains.[29] As another possibility, high-field MR in metallic NPs can be interpreted as a signature of the presence of magnetic impurities localised within the insulator barrier.[12,13,17,23] These two hypothesis are compatible with the main assumptions of our model: in fact, no clear discrimination is done in the calculations concerning the position of the localised states (at the surface of the particle, or within the insulating barrier). In both cases, paramagnetic behaviour is the clue of the high-field behaviour.

We now discuss the value of the fitting parameters. Best fit was obtained with $\mu_j = g.S.\mu_B = 1\ \mu_B$, which corresponds to a spin value of $S = 1/2$ with a Lande's factor $g = 2$. It is lower than the value expected for an isolated atom ($S = 1$ or $3/2$ in the case of Co or Fe). However, since NPs are reduced from organometallic precursors in solution in the presence of carboxylic acids and amines, the presence of infinitesimal residues of Co or Fe ions precursors cannot be excluded. If these species are present during the reaction of synthesis, carboxylate ions involving ions such as $Co^{2+}$ could be formed and may lead to $S = 1/2$ in low spin configuration depending on the ligands symmetry. The thickness ($s = 0.5$ nm) and the height ($J = 2$ V) extracted from the fits are rather common parameters for tunnel barriers. The thickness corresponds to a distance of 2 atomic layers from the surface of the particle, which could reinforce the hypothesis that the origin of the high-field effect can be ascribed to some weakly coupled impurity close to the surface of the NPs.

## VI. CONCLUSION

In conclusion, we have presented a simple model of transport through the effective tunnel barrier creating by paramagnetic impurities present at the surface of ferromagnetic nanoparticles or inside the tunnel barrier. The model takes into account the height and the thickness of the barrier, the magnetic moment of the impurity and the influence of bias voltage on MR ratio. It is shown that strong values of MR can be explained, especially for bias voltage close to the height of the barrier. Experimental behaviour of MR is well reproduced by our numerical results using Fowler-Nordheim's equation, especially the temperature and the strong bias voltage dependence of the MR amplitude.

**FIGURES CAPTIONS**

Figure 1: a) Magnetoresistance curves measured at various temperatures (between 2 K and 10 K) for $V = 200$ V, plotted as a function of the magnetic field on the left part, and as a function of the ratio $H/T$ on the right side. b) MR variation at 2.7 T as a function of temperature for $V = 200$ V. c) MR curves at $T = 3.15$ K for various bias voltages. d) $I(V)$ characteristics at $T = 2.75$ K, without magnetic field and for $\mu_0 H = 8.8$ T. Left axis represents MR $(V)$ deduced from the two $I(V)$ characteristics.

Figure 2 : Schematic illustration of the model used in the case of (A) weak magnetic field and (B) strong magnetic field.

Figure 3 : $I(V)$ characteristics calculated using a) Simmons model b) Fowler-Nordheim's model. c) MR$(V)$ curves calculated using the two models for $s = 1, 1.5, 2$ nm and $J = 1$V. d) MR$(V)$ curves calculated in both cases for $s = 1, 1.5, 2$ nm and $J = 2$ V. Vertical dash lines represent values of $\phi$ (5T) and $J$ which indicate the domain of validity of the two models (see text)

Figure 4 : a) $R(H)$ characteristics calculated at various voltage for $\mu/T = 0.3$ $\mu_B.K^{-1}$, $J = 1$ V, $s = 1$ nm using Simmons model with $R_S = 0$ $\Omega$. b) same calculations using Fowler-Nordheim's model with $R_S = 0$ $\Omega$. c) same calculations for Fowler-Nordheim with $R_S = 1$ M$\Omega$

Figure 5 : MR amplitude at 5 T a) using Simmons model and $\mu/T = 0.1$ $\mu_B.K^{-1}$ b) using Fowler-Nordheim's model and $\mu/T = 0.1$ $\mu_B.K^{-1}$ c) using Simmons model and $\mu/T = 0.3$ $\mu_B.K^{-1}$ d) using Fowler-Nordheim's model and $\mu/T = 0.3$ $\mu_B.K^{-1}$

Figure 6 : Numerical results using $\mu = 1$ $\mu_B$, $s = 0.5$ nm, $J = 2$ V, $R_S = 15$ k$\Omega$. a) MR$(T)$ variation at $V = 200$V. b) c) d) fit of the experimental $R(H)$ characteristics obtained at voltage b) $V = 200$ c) $V = 70$ V d) $V = 20$ V

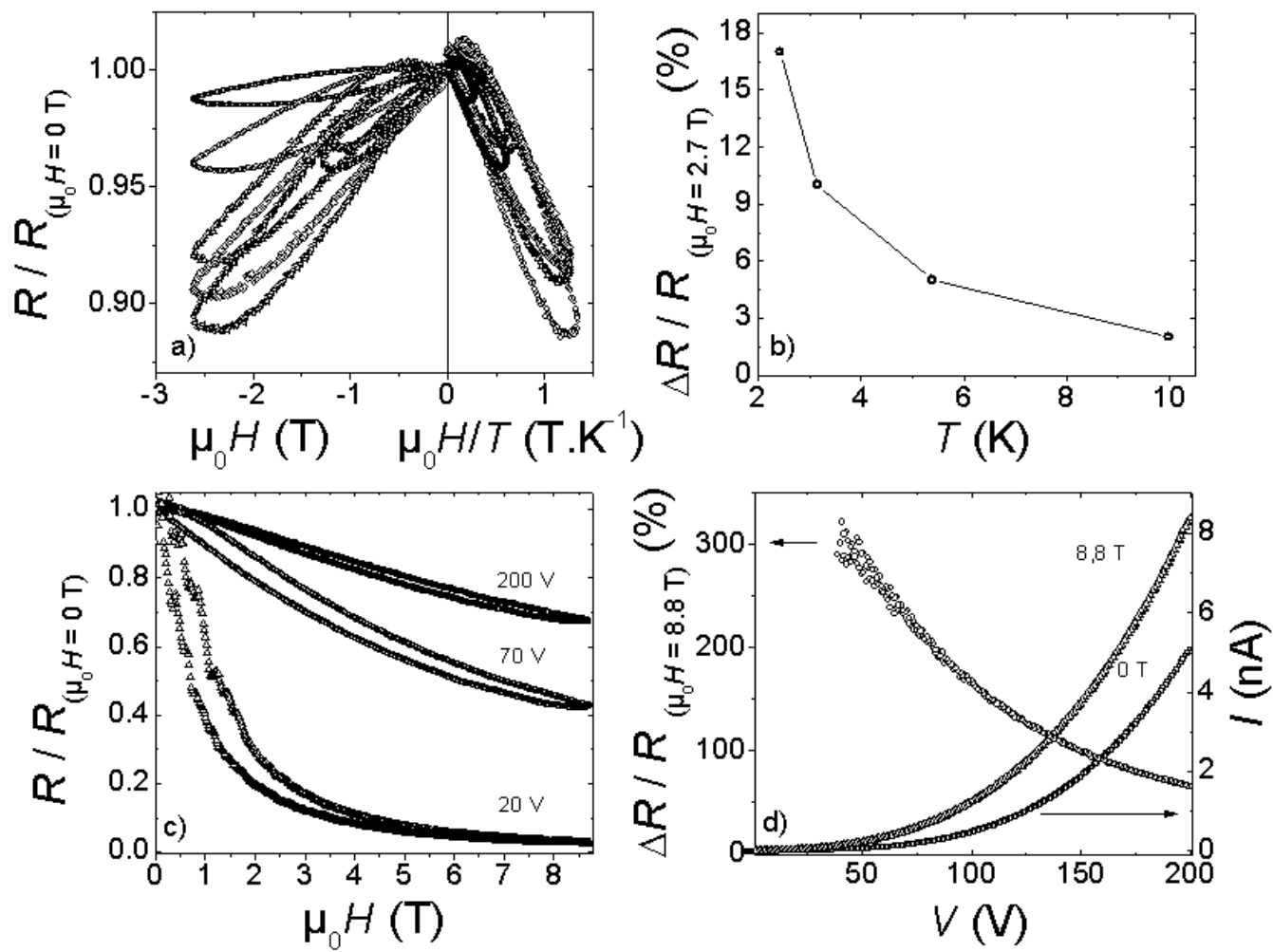

Figure 1

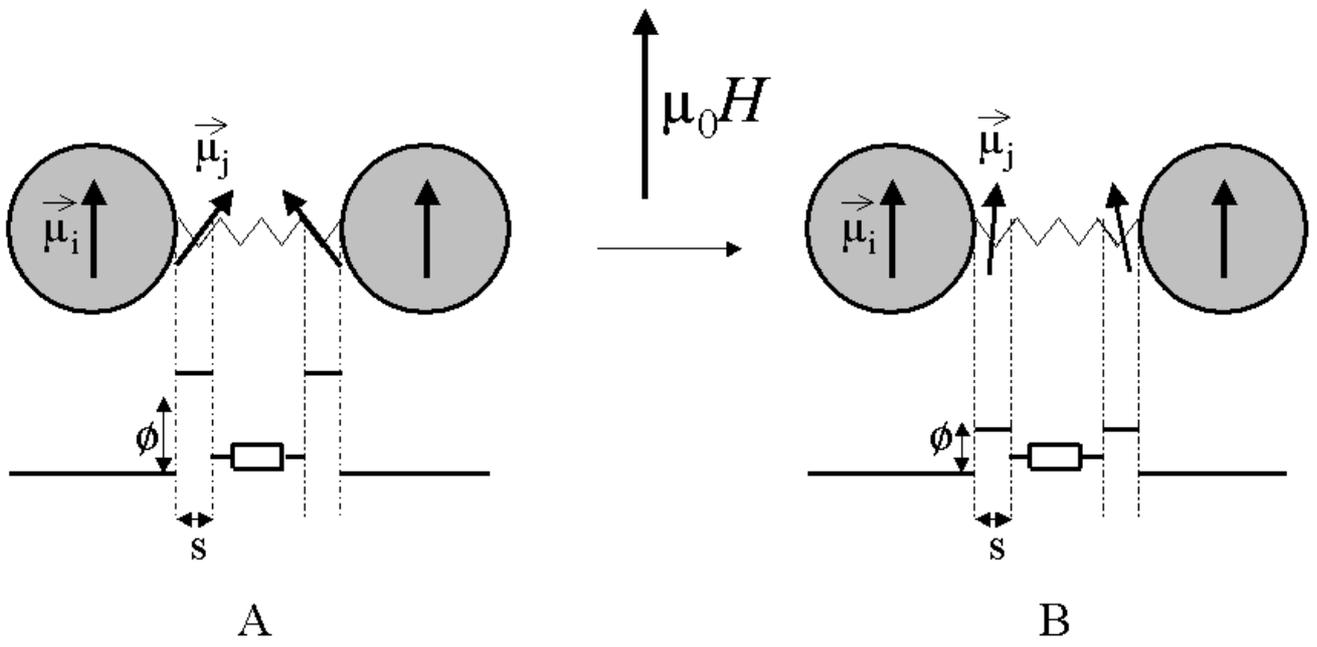

Figure 2

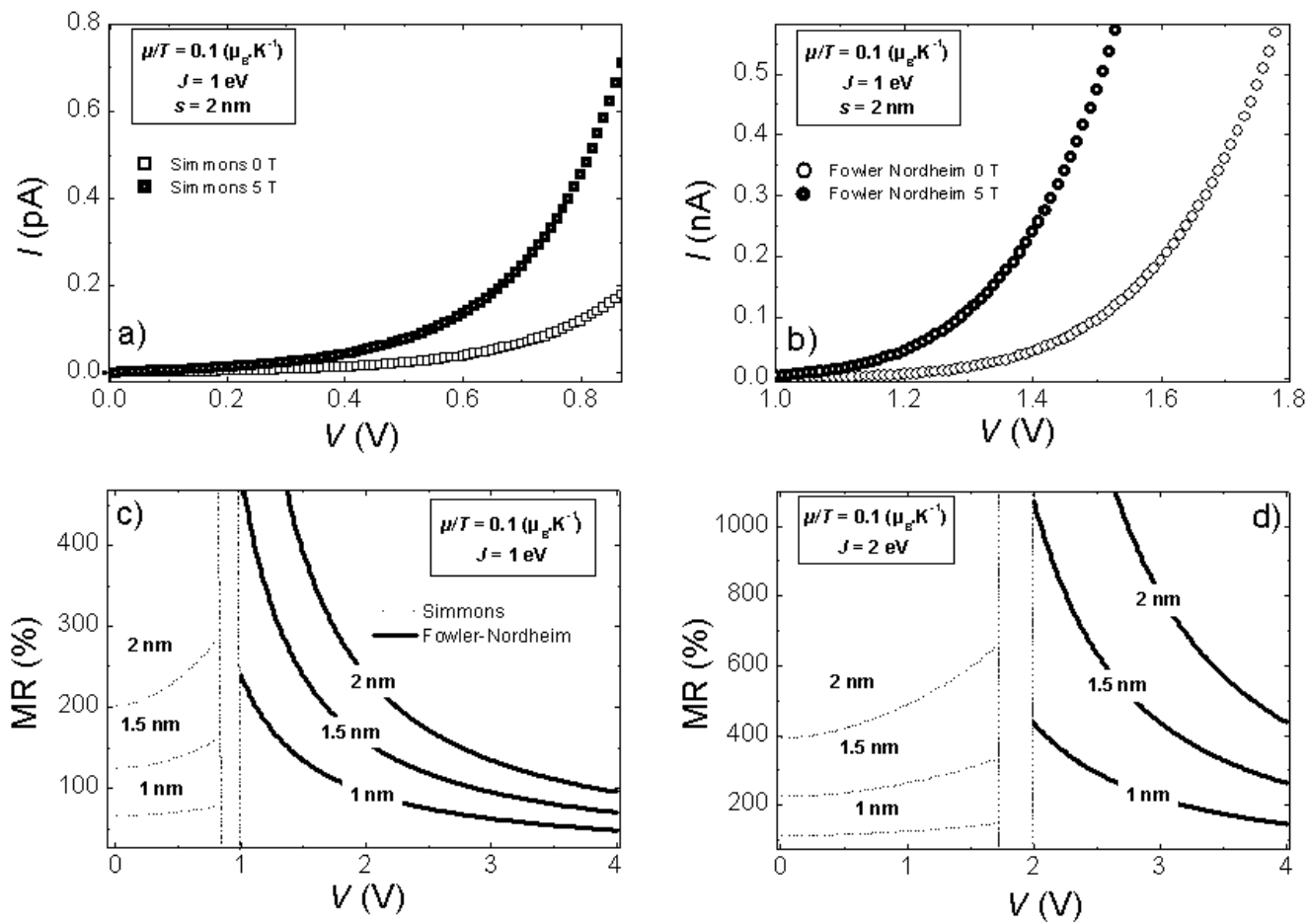

Figure 3

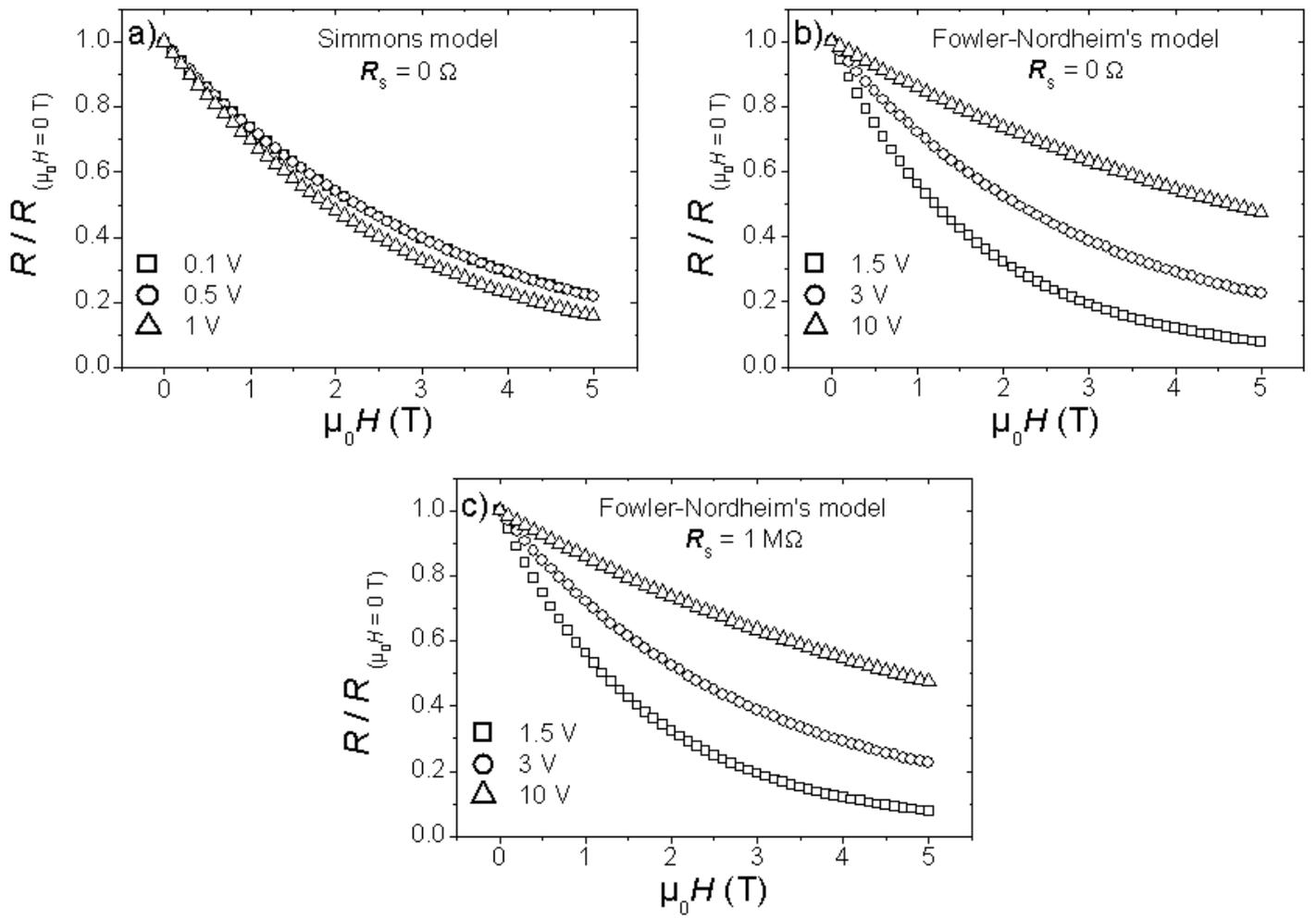

Figure 4

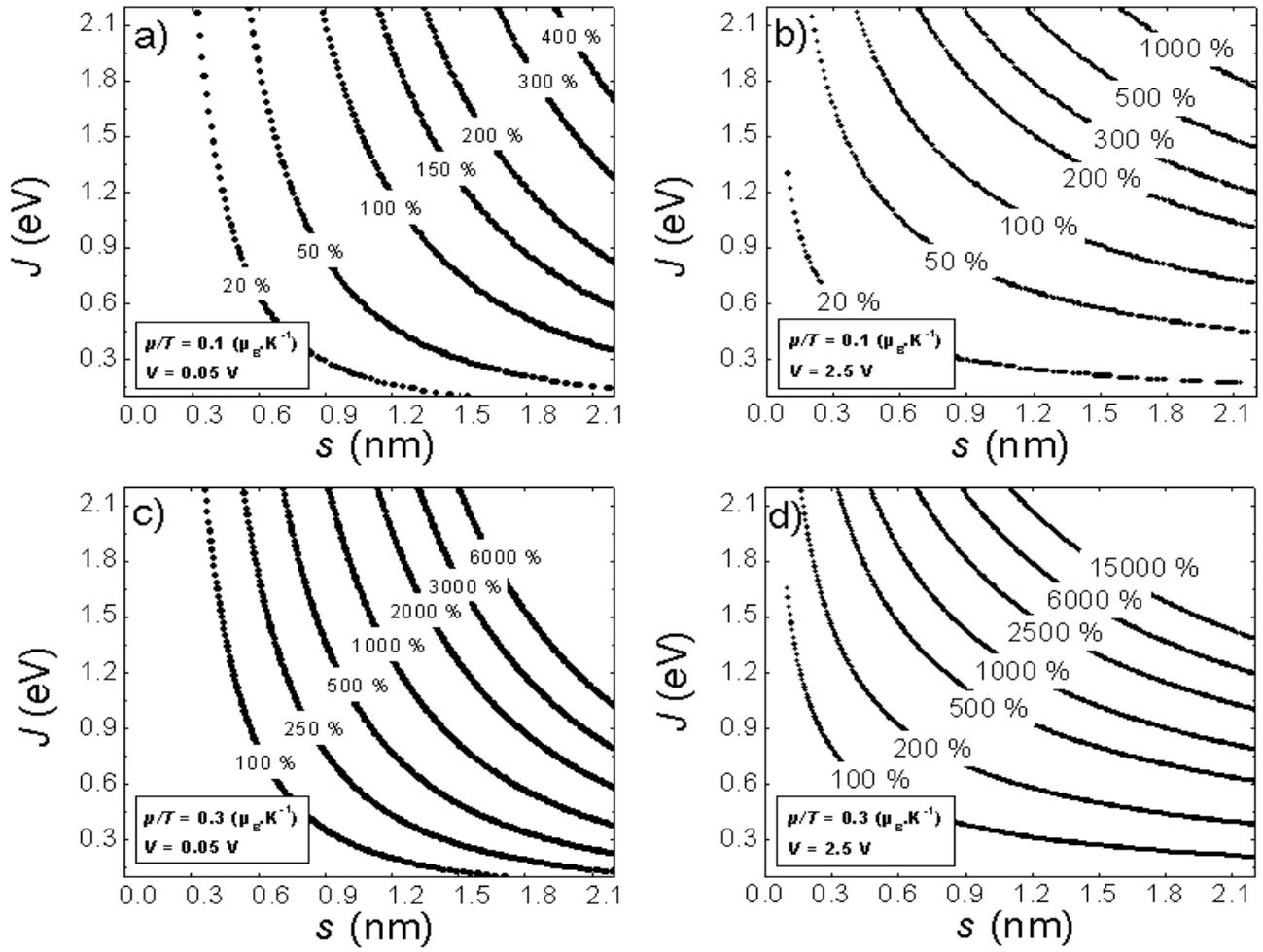

Figure 5

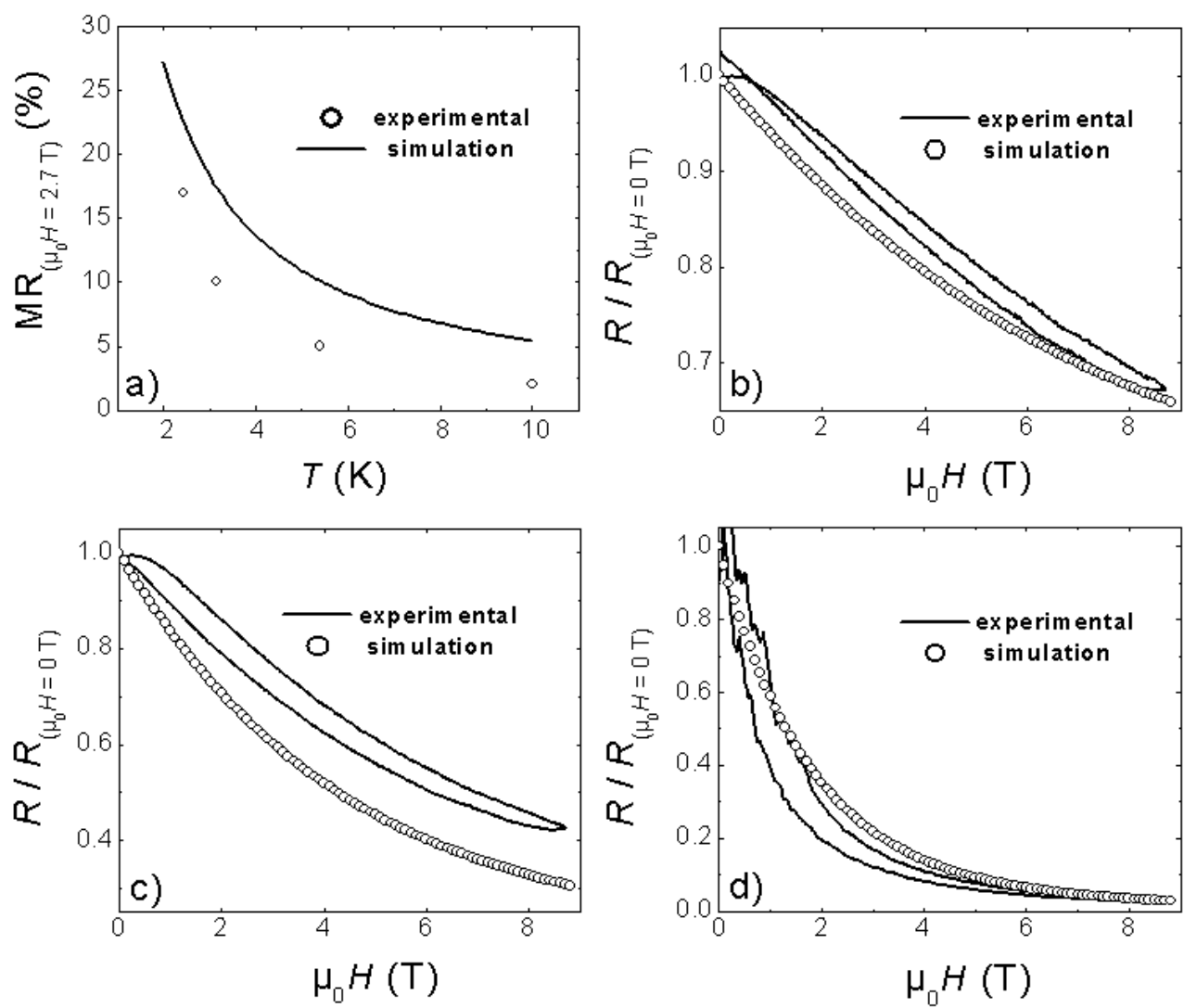

Figure 6